\documentclass[useAMS,usegraphicx]{mn2e}

\newcommand{\Msun}{\ensuremath{\,{\rm M}_\odot}}            
\newcommand{\msun}{\ensuremath{{\rm M}_\odot}}            
\newcommand{\Rsun}{\ensuremath{\,{\rm R}_\odot}}            
\newcommand{\Teff}{\ensuremath{T_{\rm eff}}}                
\newcommand{\logg}{\ensuremath{\log g}}                     
\newcommand{\Vsys}{\ensuremath{V_\gamma}}                   
\newcommand{\Vmicro}{\ensuremath{V_{\rm micro}}}            
\newcommand{\EBV}{\ensuremath{E_{B-V}}}                     
\newcommand{\kms}{\,km\,s$^{-1}$}                           
\newcommand{\ion}[2]{{#1}\,{\sc {\small{#2}}}}              
\newcommand{\MoH}{\ensuremath{\left[\frac{\rm M}{\rm H}\right]}}
\newcommand{\cms}{\,cm\,s$^{-1}$}                           
\newcommand{\lsun}{\ensuremath{L_\odot}}                    

\title[Eclipsing binaries in open clusters. III. V621\,Per in $\chi$\,Persei]
      {Eclipsing binaries in open clusters. \\ III. V621\,Per in $\chi$\,Persei}

\author[J.\ Southworth, S.\ Zucker, P.\ F.\ L.\ Maxted and B.\ Smalley]
       {J.\ Southworth$^1$\thanks{E-mail: jkt@astro.keele.ac.uk (JS), Shay.Zucker@obs.unige.ch (SZ), 
        pflm@astro.keele.ac.uk (PFLM), bs@astro.keele.ac.uk (BS)}, S.\ Zucker$^2$\footnotemark[1], 
        P.\ F.\ L.\ Maxted$^1$\footnotemark[1] and B.\ Smalley$^1$\footnotemark[1]  \\
        $^1$\,Department of Physics and Chemistry, Keele University, Staffordshire, ST5 5BG, UK \\
        $^2$\,Observatoire de Gen\`eve, 51 Ch.\ des Mailletes, CH-1290 Sauverny, Switzerland}
\begin{document} \maketitle 

\begin{abstract}
V621\,Persei is a detached eclipsing binary in the open cluster $\chi$\,Persei which is composed of an early B-type giant star and a main sequence secondary component. From high-resolution spectroscopic observations and radial velocities from the literature, we determine the orbital period to be 25.5 days and the primary velocity semiamplitude to be $K = 64.5 \pm 0.4$\kms. No trace of the secondary star has been found in the spectrum. We solve the discovery light curves of this totally-eclipsing binary and find that the surface gravity of the secondary star is $\log g_{\rm B} = 4.244 \pm 0.054$. We compare the absolute masses and radii of the two stars in the mass--radius diagram, for different possible values of the primary surface gravity, to the predictions of stellar models. We find that $\log g_{\rm A} \approx 3.55$, in agreement with values found from fitting Balmer lines with synthetic profiles. The expected masses of the two stars are 12\Msun\ and 6\Msun\ and the expected radii are 10\Rsun\ and 3\Rsun. The primary component is near the blue loop stage in its evolution. 
\end{abstract}



\section{Introduction}                    \label{introduction}    

Detached eclipsing binaries are fundamental sources of data on the properties of stars (Andersen 1991). The masses and radii of two stars in a detached eclipsing binary (dEB) can be empirically determined to accuracies of better than 1\% (e.g., Torres et al.\ 2000). Accurate effective temperatures and absolute magnitudes can be found using empirical photometric calibrations, and the distances to dEBs can likewise be found to accuracies of 5\% without any use of theoretical calculations (e.g., Clausen 2004).

As the component stars of a dEB have the same age, chemical composition and distance, their physical properties provide an important test of the predictions of theoretical stellar evolutionary models. This test is aided if the stars are quite dissimilar in mass and radius, and if they have evolved away from the main sequence. Unfortunately, as the age and interior chemical abundances of stars are not, in general, observable, stellar models of any chemical composition and age can be compared to the observed data in order to find a good fit.

Eclipsing binaries in open clusters provide a way of finding the chemical composition, age or distance of two stars with accurate masses and radii, allowing a much more discriminating test of the physical ingredients of theoretical models, for example the amount of convective core overshooting and mass loss, and the accuracy of the opacities used. Alternatively, the study of a dEB in an open cluster can be used to find the chemical composition of the cluster, its age, or its distance, avoiding the use of the technique of fitting theoretical isochrones to the morphology of stars in colour-magnitude diagrams. 

Southworth, Maxted \& Smalley (2004a, hereafter Paper\,I) studied the early-type well-detached eclipsing binaries V615\,Per and V618\,Per, members of the important young open cluster h\,Persei. As the two dEBs are members of the same stellar group, this allowed the masses and radii of four stars of the same age, chemical composition and distance to be compared to the predictions of theoretical models, a situation not previously achieved. From this procedure the metal abundance of h\,Persei was found to be $Z \approx 0.01$, in conflict with previous assumptions that the cluster has an approximately solar chemical composition.

Southworth, Maxted \& Smalley (2004b, hereafter Paper\,II) studied the high-mass dEB V453\,Cyg, a member of the sparse open cluster NGC\,6871. The masses and radii of this somewhat evolved system ($\logg = 3.731, 4.005$) were determined to accuracies of 1.5\%. Although the system consists of two quite dissimilar stars (with radii of 8.55 and 5.49\Rsun), current theoretical models of the Granada (Claret 1995), Padova (Bressan et al.\ 1993), Geneva (Schaller et al.\ 1992) and Cambridge (Pols et al.\ 1998) groups showed an impressive agreement with the stellar properties of V453\,Cyg. The Granada models also agreed well with the central condensation of the primary star derived from analysis of the apsidal motion of the dEB. This good agreement suggests that we must study dEBs which are more evolved, of higher mass, or more dissimilar in order to assess the success of the different physical ingredients of stellar models. 

V621\,Per is a member of the young open cluster $\chi$\,Persei and is more evolved, and composed of more dissimilar stars, than V453\,Cyg. $\chi$\,Persei is also regarded as a physical relation of h\,Persei (Paper\,I) so a full analysis of this system would allow the simultaneous comparison of the observed masses and radii of six stars (the components of V615\,Per and V618\,Per, Paper\,I, and V621\,Per, this work) with theoretical predictions.

\subsection{V621 Persei}

V621\,Per (Table~\ref{photpartable}) was discovered to be a dEB by Krzesi\'nski \& Pigulski (1997, hereafter KP97) from approximately 1200 images, through the broad-band $B$ and $V$ filters, of the nucleus of $\chi$\,Persei. The eclipses are total, last for approximately 1.3 days, and are about 0.12\,mag deep in both $B$ and $V$. The ascending and descending branches of one eclipse were observed in $BV$ on two successive nights but the only other observations during eclipse were 102.1 days earlier, and during totality, so the period could not be determined. 

The B2 giant component of V621\,Per is one of the brightest members of $\chi$\,Persei and has been studied several times using high-resolution optical spectroscopy to determine accurate chemical abundances. Lennon, Brown \& Dufton (1988) derived a normal helium abundance but state that different lines gave different results, which they claim could be due to the high surface gravity used [$\logg = 3.6$ (\cms)]. Dufton et al.\ (1990) found a normal abundance of helium and various metals, but a deficiency of 0.4\,dex in nitrogen and aluminium. These authors may also have been the first to note that V621\,Per is a spectroscopic binary.

Vrancken et al.\ (2000) derived a precise effective temperature and surface gravity of $\Teff = 22\,500 \pm 500$\,K and $\logg = 3.40 \pm 0.05$, based on the silicon ionisation balance and direct fitting of the H$\beta$ and H$\gamma$ absorption lines using non-LTE model atmosphere calculations. They also derived a high microturbulent velocity of 9\kms\ (or 13\kms\ from the \ion{O}{ii} lines) consistent with evolution away from the main sequence. It is notable that abundance analyses generally find larger microturbulence velocities than those usually assumed (see e.g., Dufton, Durrant \& Durrant 1981). Vrancken et al.\ derived abundances of C, N, O, Mg, Al and Si comparable to other bright B\,stars in $\chi$\,Persei, but abundances of the overall sample seem to be lower than the Sun by $0.5 \pm 0.2$\,dex. Venn et al.\ (2002) derived a boron abundance using ultraviolet spectra taken with the STIS spectrograph on board the Hubble Space Telescope. They found a lower microturbulent velocity of 4\kms, as usual in the ultraviolet wavelength region, but a macroturbulent velocity of 20\kms. They also report a value of $\MoH = -0.16 \pm 0.17$\,dex from abundance analyses of the spectral lines of light metals.

\begin{table} \begin{center} 
\caption{\label{photpartable} Identifications and photometric indices for V621\,Per from various studies. All photometric parameters refer to the combined system light (although the secondary star is much fainter than the primary). Most photometric quantities have been determined many times and the quoted values have been selected as the most representative of all determinations.
\newline $^*$\,Calculated from the system magnitude in the $V$ filter, the adopted cluster distance modulus and reddening (see section~\ref{clusterinfo}) and the canonical reddening law $A_V = 3.1 \EBV$.
\newline {\bf References:} (1) Argelander (1903); (2) Oosterhoff (1937); (3) Keller et al.\ (2001); (4) Slesnick et al.\ (2002); (5) Capilla \& Fabregat (2002); (6) Two Micron All Sky Survey; (7) Crawford, Glaspey \& Perry (1970); (8) Uribe et al.\ (2002) based on proper motion and position.}
\begin{tabular}{lr@{}lr} \hline \hline 
\                             &     & V621\,Per             & Reference \\ \hline
Bonner Durchmusterung         &     & BD\,+56\degr 576      & 1         \\
Oosterhoff number             &     & Oo\,2311              & 2         \\
Keller number                 &     & KGM 43                & 3         \\
Slesnick number               &     & SHM 47                & 4         \\ \hline
$\alpha_{2000}$               &     & 02 22 09.7            & 5         \\
$\delta_{2000}$               &  +  & 57 07 02              & 5         \\ \hline
$V$                           &     & 9.400                 & 4         \\
$B-V$                         &     & 0.294                 & 4         \\
$U-B$                         & $-$ & 0.505                 & 4         \\
$J$                           &     & 8.753 $\pm$ 0.021     & 6         \\
$H$                           &     & 8.755 $\pm$ 0.026     & 6         \\
$K_{\rm s}$                   &     & 8.712 $\pm$ 0.020     & 6         \\
$b-y$                         &     & 0.282                 & 7         \\
$m_1$                         & $-$ & 0.064                 & 7         \\
$c_1$                         &     & 0.160                 & 7         \\
$\beta$                       &     & 2.621                 & 7         \\ \hline
Spectral type                 &     & B2 {\sc III}          & 7         \\ 
$M_V$                         & $-$ & 4.04 $\pm$ 0.19       & *         \\ 
Membership probability        &     & 0.94                  & 8         \\
\hline \hline \end{tabular} \end{center} \end{table}

\subsection{$\chi$ Persei}            \label{clusterinfo}

The open clusters $\chi$\,Persei (NGC\,884) and h\,Persei (NGC\,869) together form the Perseus Double Cluster. The co-evolutionary nature of h and $\chi$ Persei has been studied many times since the seminal work of Oosterhoff (1937). The results of recent photometric studies (Marco \& Bernabeu 2001; Keller et al.\ 2001; Slesnick, Hillenbrand \& Massey 2002, Capilla \& Fabregat 2002) seem to be converging to identical values of distance modulus ($11.70 \pm 0.05$\,mag) and age ($\log\tau = 7.10 \pm 0.01$ years), which implies that h and $\chi$ Persei are physically related although there do appear to be some differences in stellar content, for example the large number of Be stars in $\chi$\,Persei. The reddening of $\chi$\,Persei is $\EBV = 0.56 \pm 0.05$, but h\,Persei displays differential reddening. These issues were discussed in detail in Paper\,I.


\section{Observations}                    \label{observations}  

Spectroscopic observations were carried out in 2002 October using the 2.5\,m Isaac Newton Telescope (INT) on La Palma. The 500\,mm camera of the Intermediate Dispersion Spectrograph (IDS) was used with a holographic 2400\,{\it l}\,mm$^{-1}$ grating and an EEV 4k\,$\times$\,2k CCD. The resulting spectra have a signal to noise ratio of approximately 60 per pixel, measured in regions with no detectable spectral lines. From measurements of the full width half maximum (FWHM) of arc lines taken for wavelength calibration we estimate that the resolution is approximately 0.2\,\AA. The main spectral window chosen for observation was 4230--4500\,\AA. This contains the \ion{Mg}{ii}\ 4481\,\AA\ line which is known to be one of the best lines for radial velocity work for early-type stars (Andersen 1975; Kilian, Montenbruck \& Nissen 1991). The \ion{He}{i}\ 4471\,\AA\ and H$\gamma$ (4340\,\AA) lines are useful for determination of effective temperatures and spectral types for such stars. One spectrum was observed at H$\beta$ (4861\,\AA) to provide an additional indicator of effective temperature. 

Data reduction was undertaken using optimal extraction as implemented in the software tools {\sc pamela} and {\sc molly}\footnote{{\sc pamela} and {\sc molly} were written by Dr.\ Tom Marsh and are found at \texttt{http://www.astro.soton.ac.uk/$^{\sim}\!$trm/software.html}} (Marsh 1989).


\section{Spectroscopic orbit}            \label{specorbit}

The INT spectra contain identifiable spectral lines from only the primary star. Radial velocities were derived from the spectra by cross-correlation with a synthetic template spectrum, using the {\sc xcor} routine in {\sc molly}. Several template spectra were investigated and the resulting radial velocities were found to be insensitive to the choice of template. 

As our spectroscopic observations cover less than the full orbital period of V621\,Per, and the orbit is eccentric, they cannot provide a unique value of the period. To solve this problem, literature radial velocities were taken from Liu, Janes \& Bania (1989, 1991) and Venn et al.\ (2002). Additional high-resolution spectra were generously made available by Dr.\ P.\ Dufton and Dr.\ D.\ Lennon. These were originally observed to determine chemical abundances (Dufton et al.\ 1990; Vrancken et al.\ 2000) so the wavelength calibrations may only be accurate to a few \kms\ (P.\ Dufton, private communication). Radial velocities were derived by fitting Gaussian functions to strong spectral lines, predominantly from the \ion{He}{i} and \ion{O}{ii} ions, and excellent agreement was found between different lines in the same spectra.

Using the photometric constraint that the orbital period of V621\,Per must be a submultiple of 102.1\,days (KP97), the possible periods were investigated by analysing the radial velocities with {\sc sbop}\footnote{Spectroscopic Binary Orbit Program written by Dr.\ Paul B.\ Etzel (\texttt{http://mintaka.sdsu.edu/faculty/etzel/}).} and the KP97 light curves with {\sc ebop} (Section~\ref{ebop}). Only a period around 25.5 days can provide a good fit to all the data. The photometric and spectroscopic data were fitted simultaneously by requiring the spectroscopically-derived orbital period to correctly predict the phase at which the primary eclipse occurs. The resulting orbital ephemeris is 
\[ T_{\rm peri} = {\rm HJD}\ 2\,452\,565.150(97) + 25.53018(20) \times E \] 
where $T_{\rm peri}$ is the time of periastron passage and the numbers in parentheses refer to the uncertainty in the last digit of the quantity. The midpoints of the primary and secondary eclipses occur at phases 0.67 and 0.06, respectively.

\begin{figure} \includegraphics[width=0.48\textwidth,angle=0]{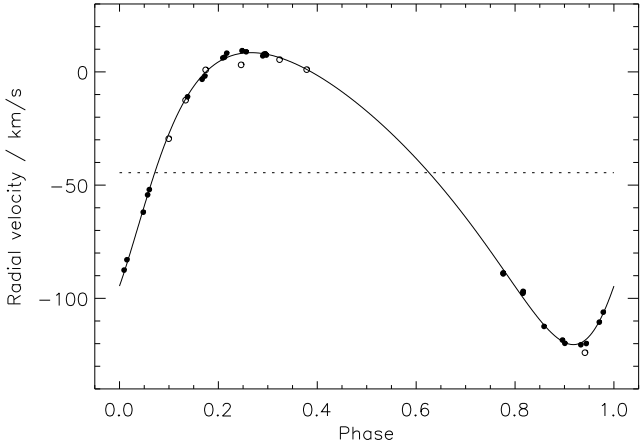} \\ \caption{\label{orbitplot} Spectroscopic orbit for V621\,Per. Filled circles denote radial velocities derived from the INT spectra and open circles show radial velocities obtained from other sources. The systemic velocity is indicated by a dotted line.} \end{figure}

The final spectroscopic orbit was calculated using the radial velocities derived from the INT spectra and fixing the orbital period at the value given above. The orbit is plotted in Fig.~\ref{orbitplot}, the radial velocities and observed minus calculated ($O-C$) values are given in Table~\ref{RVtable} and the parameters of the orbit are given in Table~\ref{spectable}.

\begin{table} \begin{center} 
\caption{\label{RVtable}Radial velocity observations of V621\,Per and the $O-C$ values with respect to the final spectroscopic orbit. 
\newline {\bf References:} (1) Liu, Janes \& Bania (1989); (2) Liu, Janes \& Bania (1991); (3) measured from spectra generously provided by Dr.\ P.\ Dufton; (4) Venn et al.\ (2002); (5) This work.}
\begin{tabular}{lrrr} \hline \hline 
HJD $-$      & Radial velocity & $O-C$     & Reference \\ 
2\,400\,000  & (\kms)          & (\kms)    &           \\ \hline
47439.8623   &      3.1   & $-$5.1   & 1 \\ 
47824.8013   &      5.4   & $-$0.9   & 2 \\   
49678.742    & $-$124.0   & $-$5.5   & 3 \\   
49682.772    &  $-$29.5   & $-$2.0   & 3 \\   
49683.643    &  $-$12.5   & $-$0.5   & 3 \\   
49684.665    &      0.9   &    1.0   & 3 \\   
51221.71538  &      1.0   & $-$0.1   & 4 \\ \hline
22559.42678  &  $-$89.0   & $-$0.1   & 5 \\   
52559.43050  &  $-$88.8   & $-$1.8   & 5 \\   
52559.43423  &  $-$89.2   & $-$1.6   & 5 \\   
52560.45020  &  $-$97.8   &    2.0   & 5 \\   
52560.45390  &  $-$97.4   &    2.3   & 5 \\   
52560.45761  &  $-$96.9   &    2.8   & 5 \\   
52561.54122  & $-$112.4   & $-$0.7   & 5 \\   
52562.49053  & $-$118.4   &    0.6   & 5 \\   
52562.61487  & $-$119.9   & $-$0.3   & 5 \\   
52563.43990  & $-$120.5   & $-$0.8   & 5 \\   
52563.71957  & $-$120.0   & $-$1.9   & 5 \\   
52564.39418  & $-$110.5   & $-$0.4   & 5 \\   
52564.59844  & $-$106.1   &    0.4   & 5 \\   
52565.38518  &  $-$87.5   &    1.1   & 5 \\   
52565.53126  &  $-$83.0   &    1.7   & 5 \\   
52566.37575  &  $-$62.0   & $-$1.2   & 5 \\   
52566.59947  &  $-$54.3   &    0.1   & 5 \\   
52566.68738  &  $-$52.0   &    0.1   & 5 \\   
52568.66096  &  $-$11.0   & $-$0.5   & 5 \\   
52569.41556  &   $-$3.3   & $-$1.5   & 5 \\   
52569.55706  &   $-$1.9   & $-$1.4   & 5 \\   
52570.48284  &      6.2   &    0.7   & 5 \\   
52570.57305  &      6.5   &    0.6   & 5 \\   
52570.68548  &      8.3   &    1.9   & 5 \\   
52571.47473  &      9.4   &    1.1   & 5 \\   
52571.68893  &      8.9   &    0.5   & 5 \\   
52572.54620  &      7.1   & $-$1.0   & 5 \\   
52572.59940  &      7.8   & $-$0.2   & 5 \\   
52572.67052  &      8.0   &    0.1   & 5 \\   
52572.73174  &      7.4   & $-$0.4   & 5 \\   
\hline \hline \end{tabular} \end{center} \end{table}

\begin{table} \begin{center} \caption{\label{spectable} Parameters of the spectroscopic orbit derived for V621\,Per.}
\begin{tabular}{l l r@{\,$\pm$\,}l} \hline \hline
Orbital period (days)         & $P$             &        25.53018 & 0.00020   \\
Reference time (HJD)          & $T_{\rm peri}$  & 2\,452\,565.150 & 0.097     \\
Eccentricity                  & $e$             &          0.2964 & 0.0057    \\
Periastron longitude (\degr)  & $\omega$        &           233.2 & 2.0       \\
Semiamplitude (\kms)          & $K$             &           64.46 & 0.40      \\
Systemic velocity (\kms)      & \Vsys           &        $-$44.53 & 0.46      \\
Mass function (\msun)         & $f(M)$          &           0.617 & 0.012     \\
\hline \hline \end{tabular}\end{center} \end{table}

The projected rotational velocity of the primary component of V621\,Per was found by fitting Gaussian functions to the \ion{Si}{iii} 4575\,\AA\ spectral line singlet. Using the orbital inclination found in Section~\ref{ebop}, the equatorial rotational velocity is $V_{\rm eq} = 32.2 \pm 1.2$\kms\ where the quoted error is the 1\,$\sigma$ error of the individual values.

The INT spectra are single-lined in character, and in experiments with the two-dimensional cross-correlation algorithm {\sc todcor} (Zucker \& Mazeh 1994) and with spectral disentangling (Simon \& Sturm 1994) we were unable to detect any signal from the secondary star. By simulating the spectrum of the secondary star with a rotationally broadened primary spectrum we have constructed several trial composite spectra. From analysis of these using cross-correlation, we estimate that we would have detected secondary spectral lines if it contributed more than 5\% of the total light, for a rotational velocity of 50\kms. If it rotates faster than this, or has a spectrum very different to that of the primary star, then the detection threshold will increase.


\section{Determination of effective temperature and surface gravity}              \label{specsynth}  

Lennon, Brown \& Dufton (1988) found the atmospheric parameters of V621\,Per to be $\Teff = 21\,500$\,K and $\logg = 3.6$, from several different Str\"omgren photometric calibrations and fitting Balmer lines with synthetic profiles. Dufton et al.\ (1990) found $\Teff = 21\,700$\,K and $\logg = 3.6$ using a similar method, but found $\Teff = 23\,000$\,K from the silicon ionization equilibrium, for which the corresponding $\logg$ is 3.7. 

Vrancken et al.\ (2000) derived $\Teff = 22\,500 \pm 500$\,K and $\logg = 3.40 \pm 0.05$ from the silicon ionization equilibrium and fitting Balmer lines with synthetic profiles. These atmospheric parameters were adopted by Venn et al.\ (2002).

Using Str\"omgren $uvby\beta$ data taken from Crawford, Glaspey \& Perry (1970), and the calibration of Moon \& Dworetsky (1985), we derive $\Teff = 21\,700 \pm 800$\,K and $\logg = 3.69 \pm 0.07$ (where the uncertainty is a formal error of the fit). The Str\"omgren photometry of Marco \& Bernabeu (2001) gives $\Teff = 19\,300 \pm 800$\,K and $\logg = 3.56 \pm 0.07$, and of Capilla \& Fabregat gives $\Teff = 20\,900 \pm 800$\,K and $\logg = 3.36 \pm 0.07$. The Crawford et al.\ (1970) photometry should be preferred as the filters are closest to the filters used to define the Str\"omgren $uvby$ and Crawford $\beta$ systems, and because photoelectric $uvby\beta$ photometry has been shown to be superior to CCD $uvby\beta$ photometry (see e.g., Mermilliod \& Paunzen 2003).

Using Geneva photometry from Rufener (1976) and the calibration of Kunzli et al.\ (1997) we find $\Teff = 22\,230 \pm 250$\,K and a high $\logg$ value of $3.97 \pm 0.18$. Use of the Geneva photometry of Waelkens et al.\ (1990) gives $\Teff = 23\,200 \pm 420$\,K and a low $\logg$ value of $3.31 \pm 0.26$.

We have assumed a surface gravity value of $\logg = 3.6$, which agrees well with Dufton et al.\ (1990), the calibration results using the Crawford et al.\ (1970) data, and with our spectroscopic and photometric analyses (see Section~\ref{absdim}). We have fitted our H$\gamma$ and H$\beta$ spectra with synthetic profiles calculated using {\sc uclsyn} (Smith 1992, Smalley et al.\ 2001; see Paper\,I for further references) and Kurucz (1993) {\sc atlas9} model atmospheres. The spectra were rotationally broadened as necessary and instrumental broadening was applied with FWHM\,=\,0.2\,\AA\ to match the resolution of the observations. For $\logg = 3.60$ we find $\Teff = 22\,500 \pm 500$\,K, in agreement with Vrancken et al.\ (2000).


\section{Light curve analysis}            \label{ebop}

\begin{table*} \begin{center} 
\caption{\label{LCtable} Results of the light curve analysis of V621\,Per for several different (fixed) values of the surface brightness ratio and orbital inclination. The final entry gives the adopted values and uncertainties of the parameters (see text for discussion). The uncertainties are confidence intervals, not 1\,$\sigma$ errors.}
\begin{tabular}{ccccccc} \hline \hline 
Light & Inclination & Surface bright- & Primary         & Secondary       & Light     & $\sigma$ ($m$mag)     \\
curve & $i$ (\degr) & ness ratio $J$  & radius ($a$)    & radius ($a$)    & ratio     & (one observation)     \\ \hline
$B$     & 90.0 & 0.0  & 0.09754 $\pm$ 0.00038 & 0.02944 $\pm$ 0.00008 &         0.0         & 4.82  \\
$B$     & 90.0 & 0.5  & 0.09753 $\pm$ 0.00039 & 0.03019 $\pm$ 0.00009 & 0.0477 $\pm$ 0.0006 & 4.76  \\
$B$     & 90.0 & 1.0  & 0.09751 $\pm$ 0.00039 & 0.03098 $\pm$ 0.00009 & 0.1005 $\pm$ 0.0014 & 4.71  \\[2pt]
$B$     & 89.0 & 0.0  & 0.09951 $\pm$ 0.00037 & 0.03020 $\pm$ 0.00008 &         0.0         & 4.79  \\
$B$     & 89.0 & 0.5  & 0.09950 $\pm$ 0.00038 & 0.03096 $\pm$ 0.00009 & 0.0482 $\pm$ 0.0006 & 4.75  \\
$B$     & 89.0 & 1.0  & 0.09948 $\pm$ 0.00038 & 0.03178 $\pm$ 0.00009 & 0.1016 $\pm$ 0.     & 4.71  \\[2pt]
$B$     & 88.0 & 0.0  & 0.10517 $\pm$ 0.00036 & 0.03236 $\pm$ 0.00008 &         0.0         & 4.91  \\
$B$     & 88.0 & 0.5  & 0.10515 $\pm$ 0.00037 & 0.03320 $\pm$ 0.00009 & 0.0496 $\pm$ 0.0007 & 4.90  \\
$B$     & 88.0 & 1.0  & 0.10514 $\pm$ 0.00037 & 0.03411 $\pm$ 0.00010 & 0.1046 $\pm$ 0.0016 & 4.91  \\ \hline
$V$     & 90.0 & 0.0  & 0.09792 $\pm$ 0.00032 & 0.02984 $\pm$ 0.00007 &         0.0         & 4.83  \\
$V$     & 90.0 & 0.5  & 0.09791 $\pm$ 0.00032 & 0.03062 $\pm$ 0.00007 & 0.0487 $\pm$ 0.0007 & 4.78  \\
$V$     & 90.0 & 1.0  & 0.09790 $\pm$ 0.00216 & 0.03147 $\pm$ 0.00205 & 0.1028 $\pm$ 0.2972 & 4.74  \\[2pt]
$V$     & 89.0 & 0.0  & 0.09995 $\pm$ 0.00033 & 0.03066 $\pm$ 0.00006 &         0.0         & 4.73  \\
$V$     & 89.0 & 0.5  & 0.09991 $\pm$ 0.00112 & 0.03147 $\pm$ 0.00283 & 0.0494 $\pm$ 0.0994 & 4.71  \\
$V$     & 89.0 & 1.0  & 0.09992 $\pm$ 0.00158 & 0.03236 $\pm$ 0.00147 & 0.1044 $\pm$ 0.2070 & 4.71  \\[2pt]
$V$     & 88.0 & 0.0  & 0.10580 $\pm$ 0.00031 & 0.03307 $\pm$ 0.00007 &         0.0         & 4.83  \\
$V$     & 88.0 & 0.5  & 0.10579 $\pm$ 0.00122 & 0.03398 $\pm$ 0.00281 & 0.0513 $\pm$ 0.0960 & 4.91  \\
$V$     & 88.0 & 1.0  & 0.10579 $\pm$ 0.00032 & 0.03498 $\pm$ 0.00009 & 0.1087 $\pm$ 0.0017 & 5.02  \\ \hline
Adopted & 89.0 $\pm$ 1.0 & 0.25 $\pm$ 0.25  &   0.1016 $\pm$ 0.0.0039 & 0.0316 $\pm$ 0.0020 &       \\
\hline \hline \end{tabular} \end{center} \end{table*}

\begin{figure*} \includegraphics[width=\textwidth,angle=0]{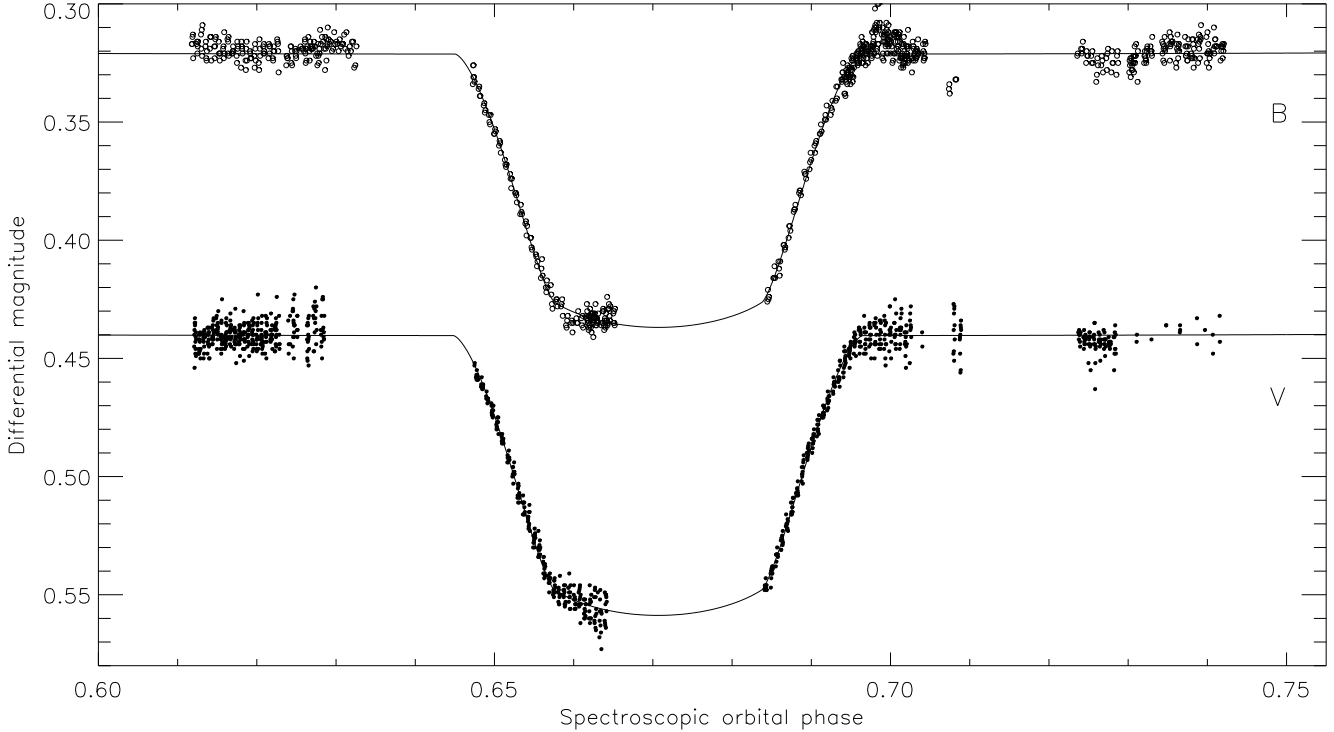} \\ \caption{\label{lcfitfigure} 
The KP97 $B$ and $V$ light curves of V621\,Per around the primary eclipse, phased using the spectroscopic ephemeris, with the best-fitting {\sc ebop} model light curves. The $B$ light curve, offset by $-0.1$\,mag for clarity, is shown using open circles and the $V$ light curve is shown using filled circles. The best-fitting curves were generated with $J = 0.25$ and $i = 89.0$\degr.} \end{figure*}

\begin{figure} \includegraphics[width=0.48\textwidth,angle=0]{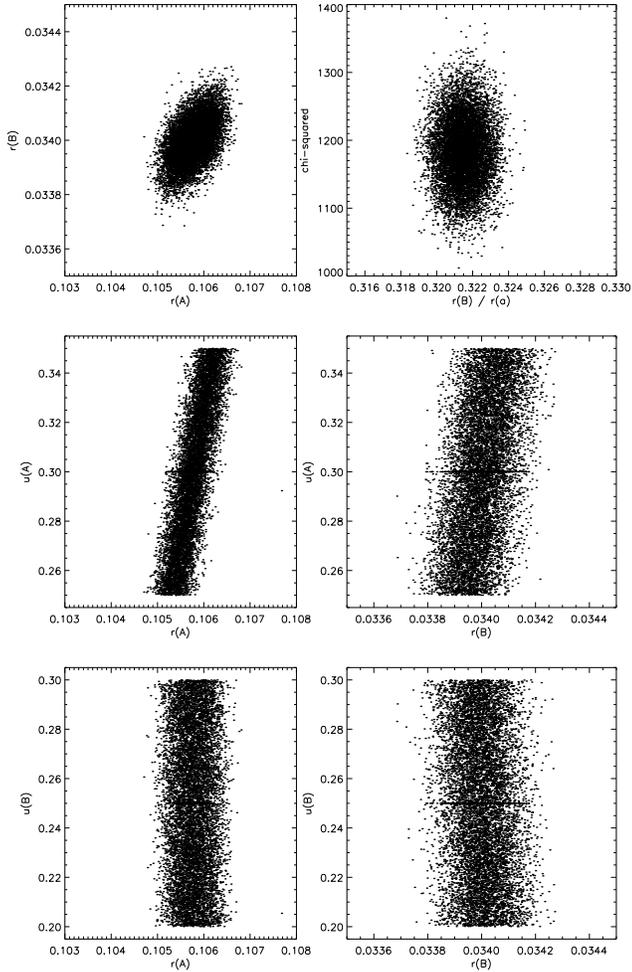} \\ \caption{\label{lcerrs} 
Results of the Monte Carlo analysis for $J = 0.25$ and $i = 89.0$\degr. The limb darkening coefficients, $u_{\rm A}$ and $u_{\rm B}$, were chosen randomly on a flat distribution between $u - 0.05$ and $u + 0.05$ for each synthetic light curve, and fixed during solution of the light curve.} \end{figure}

We have analysed the $BV$ light curves of KP97 using the simple and efficient NDE eclipsing binary model (Nelson \& Davis 1972) as implemented in the light curve analysis program {\sc ebop}\footnote{Eclipsing Binary Orbit Program written by Dr.\ Paul B.\ Etzel (\texttt{http://mintaka.sdsu.edu/faculty/etzel/}).} (Popper \& Etzel 1981). In this model the stellar shapes are approximated by biaxial ellipsoids. The orbital eccentricity and longitude of periastron were fixed at the spectroscopic values, initial filter-specific linear limb darkening coefficients of 0.30 (primary star) and 0.25 (secondary star) were taken from van Hamme (1993) and the gravity darkening exponents $\beta_1$ were fixed at 1.0 (Claret 1998). Changes in the limb darkening and gravity brightening values for the secondary star have a negligible effect on the photometric solutions because this star contributes very little of the light of the system. Contaminating `third' light was fixed at zero, as solutions in which it was a free parameter were not significantly different from solutions with no third light.

Initial light curve solutions converged to an orbital inclination, $i$, of 90\degr, but values of $i$ from about 88 to 90\degr\ fit the observations almost equally well. Solutions in which the surface brightness ratio, $J$, was freely adjusted towards the best fit generally converged to a value of $J$ below zero, which is unphysical. We therefore present separate solutions (Table~\ref{LCtable}) for the $B$ and $V$ light curves in which orbital inclination is fixed at $i = 88, 89$ and 90\degr\ and the surface brightness ratio is fixed at $J = 0.0, 0.5$ and 1.0. The light curves are of insufficient quality to solve for the limb darkening coefficients so these have been fixed during solution.

Robust errors were estimated using Monto Carlo simulations (see Paper\,II for a detailed description of the method). In this procedure, a best fitting model is found for the observed light curve. Ten thousand synthetic light curves are then generated by evaluating the model brightness at each phase of observation, and adding observational noise of the same magnitude as that present in the observed light curve. Each synthetic light curve is fitted, and the results of the fits allow a robust estimation of the uncertainties and of the parameter correlations at the point of best fit. This method was modified to explicitly include uncertainties due to the use of assumed limb darkening coefficients by fixing them at random values on a flat distribution within $\pm$0.05 of the original value. 


As no trace of the secondary star was found in the observed spectra, the $B$-filter light ratio must be 0.05 or less. The maximum light ratio in the $V$ filter will be slightly greater than this as the secondary star is expected to have a lower effective temperature than the primary star. For simplicity, we have adopted a maximum light ratio of 0.05 for both the $B$ and $V$ light curves. We have therefore calculated best-estimate parameters by evaluating the ranges of possible parameter values in the two light curves and then averaging the midpoints of the ranges for the two light curves (Table~\ref{LCtable}). The quoted uncertainties are confidence intervals which encompass the range of possible values for each parameter, so are not 1\,$\sigma$ errors. This procedure is simple but is quite adequate considering the nature of the observations analysed here.

Fig.~\ref{lcfitfigure} shows the observed light curves and the best-fitting models with $J = 0.25$ and $i = 89$\degr. Fig.~\ref{lcerrs} represents the relation between different parameters of the fit to the $V$ light curve. As V621\,Per exhibits total eclipses, the radii of the two stars are only weakly correlated. Changes in the limb darkening coefficients used do affect the derived radii of both stars, but this effect is quite small and easily quantified.


\section{Absolute dimensions and comparison with stellar models}        \label{absdim}

Although the absolute masses and radii of the component stars of V621\,Per cannot be found directly, the mass function and fractional radii (the stellar radii expressed as a fraction of the semi-major axis of the orbit) are accurately known. This allows us to empirically determine the surface gravity of the secondary star despite not knowing its actual mass or radius.

From consideration of the definitions of the mass function, fractional radius and surface gravity, and from Kepler's third law, it can be shown that the surface gravity of the components of a single-lined dEB can be given by
\begin{equation} \label{eq:grav}
g_n = \left(\frac{2\pi}{P}\right)^{4/3} \frac{[Gf(M)]^{1/3}}{{r_n^{\ 2}\sin i}} \left(\frac{M_n}{M_{\rm B}}\right) 
\end{equation}
where $P$ is the orbital period, $f(M)$ is the mass function, $r_n$ and $M_n$ are the fractional radius and absolute mass of component $n$ ($n = {\rm A,B}$) and $M_{\rm B}$ is the mass of the spectroscopically unseen star. In the usual astrophysical units of solar masses, days and gravity in \cms, the surface gravities of the two stars are therefore given by 
\begin{equation} \label{eq:logg2}
\log g_{\rm B} = 3.18987 + \frac{\log f(M)}{3} - \frac{4\log P}{3} - \log(r_{\rm B}^{\ 2}\sin i)
\end{equation}
\begin{equation} \label{eq:logg1}
\log g_{\rm A} = 3.18987 + \frac{\log f(M)}{3} - \frac{4\log P}{3} - \log(r_{\rm A}^{\ 2}\sin i) - \log q
\end{equation}
where $q = \frac{M_{\rm B}}{M_{\rm A}}$ is the mass ratio. Equation~\ref{eq:logg2} contains only known quantities so, despite not knowing the mass {\it or} radius of the secondary component of V621\,Per we can empirically calculate its surface gravity to be $\log g_{\rm B} = 4.244 \pm 0.054$. 

Alternatively, it is possible to use V621\,Per's membership of the open cluster $\chi$\,Persei to infer the properties of the primary star. The absolute magnitudes of the V621\,Per system, found from the apparent magnitudes, the distance modulus and reddening of the cluster (Table~\ref{photpartable} and Section~\ref{clusterinfo}) and the reddening laws $A_V = 3.1 \EBV$ and $A_K = 0.38 \EBV$ (Moro \& Munari 2000), are $M_V = -4.04 \pm 0.16$ and $M_K = -3.20 \pm 0.06$. Adopting bolometric corrections of $-2.20 \pm 0.05$ and $-2.93 \pm 0.08$ (Bessell, Castelli \& Plez 1998) gives absolute bolometric magnitudes of $-6.24 \pm 0.17$ and $-6.13 \pm 0.10$ for the $V$ and $K$-filter data respectively. The two values are in good agreement but the $K$-filter value is more accurate because it is less affected by the uncertainty in \EBV. We will adopt the $K$-filter value as it is more accurate, and the 2MASS apparent magnitudes are known to be very reliable. Adopting a solar absolute bolometric magnitude of 4.74 (Bessell et al.\ 1998), which is consistent with the adopted bolometric corrections, gives a luminosity of $\log\frac{L}{\lsun} = 4.348 \pm 0.039$. This gives a radius of $9.9 \pm 0.7$\Rsun\ for the primary star. If we assume that the secondary star's contribution to the total light of the system is 5\%, this will cause the primary radius to be overestimated by about 0.25\Rsun, which is negligible at this level of accuracy.

\subsection{Comparison with stellar models}

\begin{table*} \begin{center} \caption{\label{MRtable} Absolute masses and radii of the components of V621\,Per calculated using different values of $\log g_{\rm A}$. The luminosity is that of the primary component only as we do not know the effective temperature of the secondary star.}
\begin{tabular}{ccccccc} \hline \hline
$\log g_{\rm A}$ & Mass ratio & Primary mass & Secondary mass & Primary radius & Secondary radius & Luminosity         \\
(\cms)           &            & (\Msun)      & (\Msun)        & (\Rsun)        & (\Rsun)          & ($\log L/\lsun$)   \\ \hline
3.40 & 0.676 $\pm$ 0.049 &  5.6 $\pm$  1.0 &  3.8 $\pm$  0.4 &  7.8 $\pm$  0.7 &  2.4 $\pm$  0.2 &  4.152 $\pm$  0.057 \\
3.45 & 0.602 $\pm$ 0.044 &  7.3 $\pm$  1.3 &  4.4 $\pm$  0.4 &  8.4 $\pm$  0.7 &  2.6 $\pm$  0.2 &  4.213 $\pm$  0.058 \\
3.50 & 0.537 $\pm$ 0.039 &  9.4 $\pm$  1.8 &  5.1 $\pm$  0.5 &  9.0 $\pm$  0.8 &  2.8 $\pm$  0.2 &  4.277 $\pm$  0.059 \\
3.55 & 0.478 $\pm$ 0.035 & 12.3 $\pm$  2.4 &  5.9 $\pm$  0.6 &  9.8 $\pm$  0.9 &  3.0 $\pm$  0.2 &  4.343 $\pm$  0.061 \\
3.60 & 0.426 $\pm$ 0.031 & 16.2 $\pm$  3.2 &  6.9 $\pm$  0.8 & 10.6 $\pm$  1.0 &  3.3 $\pm$  0.3 &  4.412 $\pm$  0.062 \\
3.65 & 0.380 $\pm$ 0.028 & 21.4 $\pm$  4.3 &  8.1 $\pm$  0.9 & 11.5 $\pm$  1.1 &  3.6 $\pm$  0.3 &  4.483 $\pm$  0.063 \\
3.70 & 0.339 $\pm$ 0.025 & 28.5 $\pm$  5.9 &  9.6 $\pm$  1.2 & 12.5 $\pm$  1.2 &  3.9 $\pm$  0.3 &  4.557 $\pm$  0.064 \\
\hline \hline \end{tabular}\end{center} \end{table*}

\begin{figure*} \includegraphics[width=\textwidth,angle=0]{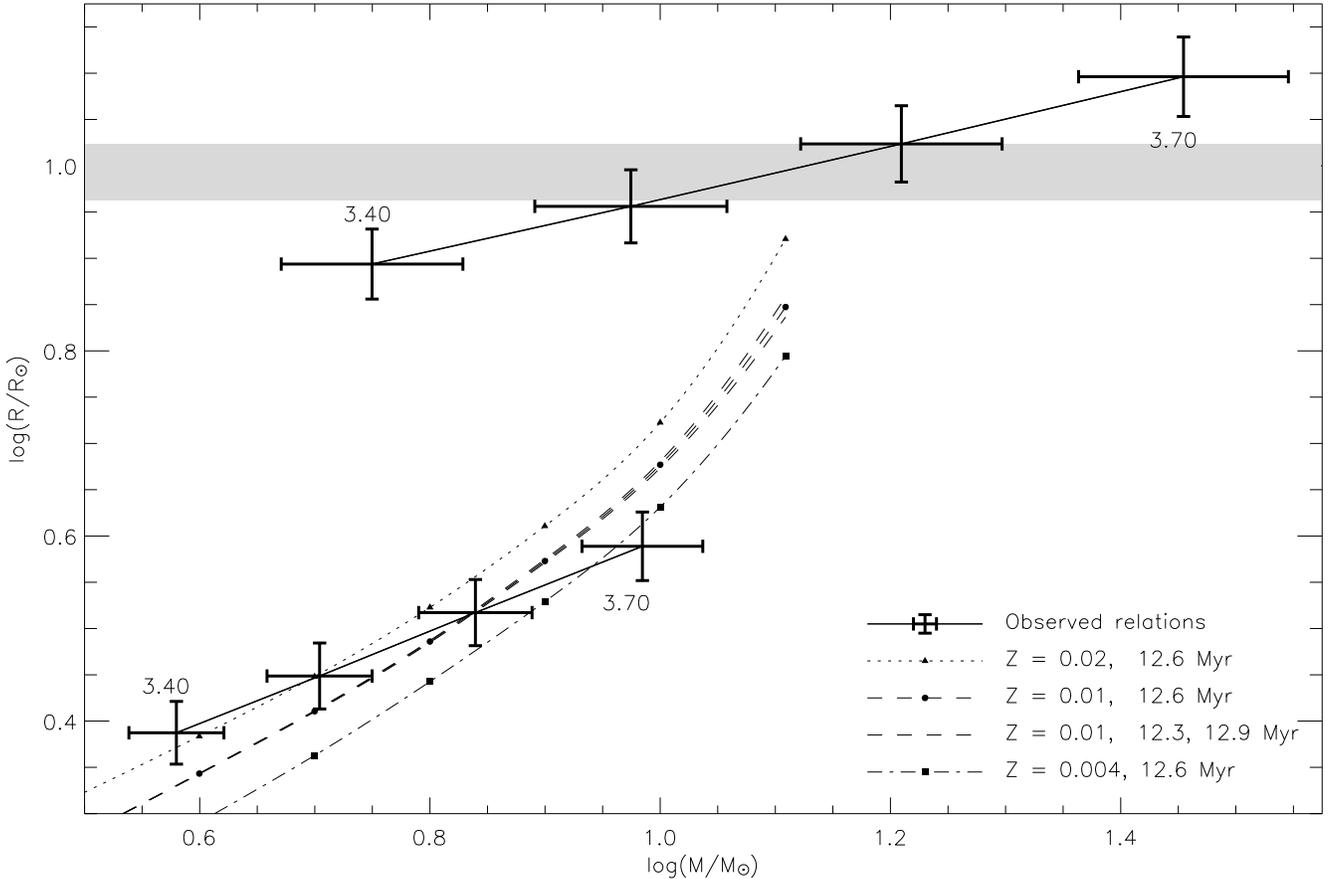} \\ \caption{\label{modelfit} The logarithmic mass--radius plot for the two components of V621\,Per. The possible combinations of mass and radius for each star, for different values of $\log g_{\rm A}$, are plotted with errorbars connected by solid lines. Numbers on the diagram indicate the value of $\log g_{\rm A}$ used to calculate the adjacent datapoint. The Granada stellar model preditions are plotted for an age of 11.0\,Myr for $Z = 0.004$ (dash-dotted line), $Z = 0.01$ (dashed line) and $Z = 0.02$ (dotted line). The 11.5\,Myr $Z = 0.01$ predictions are plotted using a dashed line with the symbols representing the points of model evaluation omitted. Curves have been calculated using a cubic spline interpolation. The radius of the primary star, calculated from its known distance and apparent magnitude, has been shown using light shading to indicate the range of possible values.} \end{figure*}

\begin{figure} \includegraphics[width=0.48\textwidth,angle=0]{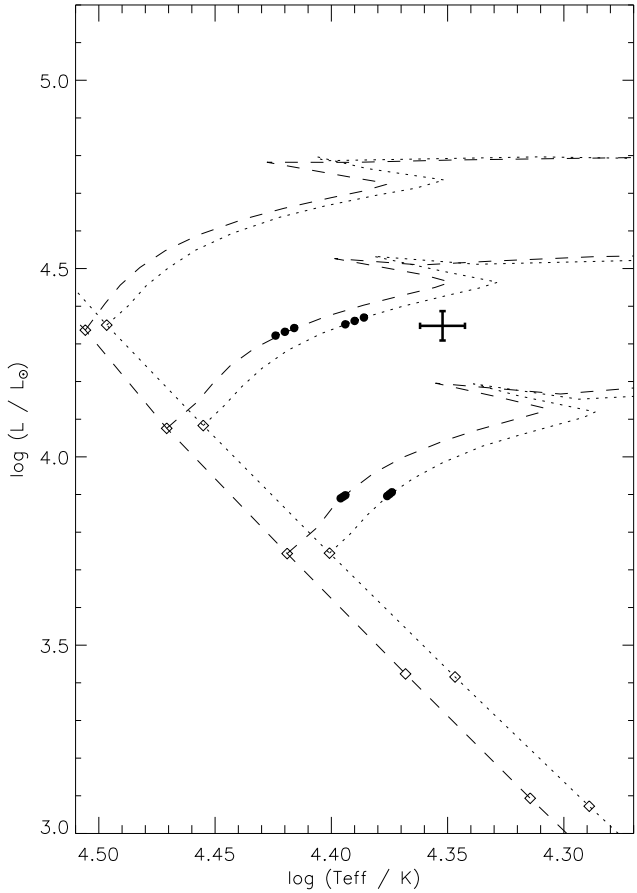} \\ \caption{\label{HRDfit} Hertzsprung-Russell diagram showing the luminosity and effective temperature derived for V621\,Per. The Granada evolutionary model preditions are plotted for for masses of 10.0, 12.8 and 15.8\Msun, and for metal abundances of $Z = 0.01$ (dashed lines) and 0.02 (dotted lines). The zero age main sequences are plotted using the same line styles and filled circles denote predictions for ages of 12.3, 12.6 and 12.9\,Myr.} \end{figure}

The mass ratio of V621\,Per can in principle be found from equation~\ref{eq:logg1} as $\log g_{\rm A}$ has been determined by several authors (see Section~\ref{specsynth}). However, the mass ratio has a very sensitive dependence on the value of $\log g_{\rm A}$ as it is effectively calculated from the difference of two numbers of similar size. An alternative approach is to evaluate the mass ratio, and hence the absolute masses and radii of the two stars, for several different values of $\log g_{\rm A}$ and compare the possibilities with the predictions of stellar models. Substituting the mass ratio into the definition of the mass function, we can derive the absolute masses of the two stars using
\begin{equation} \label{eq:m1}
M_{\rm A} = \frac{f(M)}{\sin^3i} \frac{(1+q)^2}{q^3}
\end{equation}
\begin{equation} \label{eq:m2}
M_{\rm B} = q M_{\rm A}
\end{equation}
The absolute radii of the stars can then be found from the masses and the surface gravities. 

We have used the equations above to determine the absolute masses and radii of both components of V621\,Per for several assumed values of $\log g_{\rm A}$ between 3.40 and 3.70. These have been compared to the predictions of the Granada stellar evolutionary models (Claret 1995; Claret \& Gim\'enez 1995, 1998) for ages close to the age of the $\chi$\,Persei open cluster, $12.6 \pm 0.3$\,Myr (Paper\,I). 

Fig.~\ref{modelfit} shows possible values of the absolute masses and radii of the components of V621\,Per compared to predictions of the Granada evolutionary models for ages around 12.6\,Myr. Also shown (by a shaded area) is the range of possible values of the primary radius, derived from the known distance and apparent magnitude of the dEB. Whilst all three possible metal abundances can fit the two stars for $\log g_{\rm A} \sim 3.55$, the $Z = 0.01$ predictions provide the best fit to the predicted properties of the secondary component. This diagram suggests that $\log g_{\rm A}$ is probably between 3.55 and 3.60. If the best fit is sought for metal abundances of $Z = 0.02$ and 0.004, ages of roughly 5 and 40\,Myr, respectively, are found.

The primary component of V621\,Per has been plotted in the Hertzsprung-Russell diagram (Fig.~\ref{HRDfit}) and compared with Granada theoretical model predictions for masses of 10.0, 12.8 and 15.8\Msun\ and for metal abundances of $Z = 0.01$ and 0.02. The $\chi$\,Persei open cluster has been found to have an age of $12.6 \pm 0.3$\,Myr (Paper\,I) and this age has been indicated on the evolutionary track for for each model mass. The position of the primary component of V621\,Per in these diagrams suggests that its mass is a little below 12.8\Msun, also consistent with the form of the mass--radius diagram (Fig.~\ref{modelfit}). However, its age derived by comparison with evolutionary models is somewhat greater than the 12.6\,Myr expected due to its membership of $\chi$\,Persei, and the discrepancy is larger compared to the $Z = 0.01$ model predictions than for the $Z = 0.02$ predictions. 

The age of $12.6 \pm 0.3$\,Myr for $\chi$\,Persei was derived, from comparison between photometric observations of the cluster and the predictions of theoretical models, by researchers who assumed that $Z = 0.02$ (Paper\,I). As the metal abundance of $\chi$\,Persei has been found to be $Z = 0.01$ (Paper\,I), this age may have a systematic error. Therefore the age discrepancy found here is of only minor significance, but deserves investigating when more accurate parameters are determined for V621\,Per. Also, the amount of overshooting present in stellar models is known to significantly change the predicted ages of giant stars (e.g., Schr\"oder \& Eggleton 1996) and the inclusion of rotation also affects the main sequence lifetime of high-mass stars (Maeder \& Meynet 2000). 

From comparison with the Granada evolutionary models, the surface gravity of the primary component of V621\,Per is approximately 3.55. This conclusion is valid for metal abundances of $Z = 0.01$ and 0.02. From Table~\ref{MRtable} the masses and radii of V621\,Per corresponding to $\log g_{\rm A} = 3.55$ are about 12 and 6\Msun, and 10 and 3\Rsun, for primary and secondary star respectively. This means that the primary star is near the age at which it passes through the `blue loop' evolutionary stage (the point at which core hydrogen exhaustion causes the effective temperature and surface gravity to rise temporarily), so accurate masses and radii for it would provide extremely good tests of the predictions of theoretical models. The Granada stellar models predict a luminosity ratio of about 0.05 for the inferred properties of V621\,Per. The light ratio in the blue will be smaller than the overall luminosity ratio because the secondary star has a lower effective temperature than the primary star. This suggests that the quality of our spectroscopic observations was almost sufficient to detect the secondary star. Accurate velocities for both stars should be measurable on spectra of a high signal to noise ratio, depending on the rotational velocity of the secondary star.

\subsection{Membership of the open cluster $\chi$\,Persei} 
 
V621\,Per is situated, on the sky, in the centre of the $\chi$\,Persei open cluster. It also appears in the correct place on the colour-magnitude diagram of the cluster (see Section~\ref{clusterinfo}) and has the correct proper-motion (Uribe et al.\ 2002) for cluster membership. The systemic velocity of the dEB, $-44.5 \pm 0.4$\kms, is consistent with the measured cluster systemic velocities of Oosterhoff (1937), Bidelman (1943), Hron (1987), Liu et al.\ (1989, 1991) and Chen, Hou \& Wang (2003). In Paper\,I we measured the systemic velocity of the h\,Persei cluster to be $-44.2 \pm 0.3$\kms, indicating that h and $\chi$\,Persei have the same systemic velocities, which is consistent with them having a common origin.


\section{Discussion}

V621\,Persei is a detached eclipsing binary in the young open cluster $\chi$\,Persei, composed of a bright B2 giant star and an unseen main sequence secondary star. From blue-band spectroscopic data and radial velocities taken from the literature, we have derived an orbital period of 25.5302\,days and a mass function of $f(M) = 0.617 \pm 0.012$\Msun. The discovery light curve of KP97 shows that the system exhibits a total primary eclipse lasting around 1.3\,days and about 0.12\,mag deep in $B$ and $V$. No data exist around phase 0.06, where the secondary eclipse is expected to occur. The secondary eclipse may be up to about 0.06\,mag deep. The light curves have been solved using {\sc ebop} and Monte Carlo simulations to find robust uncertainties, and accurate fractional radii have been determined. The surface gravity of the secondary component has been found to be $\log g_{\rm B} = 4.244 \pm 0.054$.

Using the data above, possible values of the absolute masses and radii of the two stars were calculated by assuming different values of the primary surface gravity. A comparison in the mass--radius diagram of these possible values with theoretical predictions from the Granada stellar evolutionary models suggests that $\log g_{\rm A} \approx 3.55$. This surface gravity value agrees well with the values determined by Lennon et al.\ (1988) and Dufton et al.\ (1990) by fitting observed Balmer line profiles with synthetic spectra. The luminosity of V621\,Per has been derived from the known distance of the $\chi$\,Persei cluster and the apparent magnitude of the dEB. This has been used to place the primary star in the Hertzsprung-Russell diagram, and a comparison with the Granada evolutionary models confirms that its mass is roughly 12\Msun, although a small discrepancy exists between the inferred age of V621\,Per and the age of $\chi$\,Persei.

The value of $\log g_{\rm A}$ leads to masses of approximately 12 and 6\Msun\ and radii of 10 and 3\Rsun\ for the components of the dEB. This conclusion is not strongly dependent on use of the Granada stellar models; predictions of the Geneva, Padova and Cambridge models are close to those of the Granada models (Paper\,II). V621\,Persei is a potentially important object for the information it holds about the evolution of high-mass stars. The expected luminosity ratio of the system, about 0.05, suggests that spectral lines of the secondary component should be detectable in high signal-to-noise spectra. Better light curves will be needed for detailed studies of the properties of V621\,Per, but the long period and lengthy eclipses mean that a large amount of telescope time will be required. In particular, the secondary eclipse is expected to occur around phase 0.06 and may be up to 0.06\,mag deep. Observations of the light variation through secondary eclipse will be needed to provide a definitive study of the system. The absolute dimensions of the primary star, a B2 giant which is close to the blue loop stage of stellar evolution, could provide a good test of the success of stellar evolutionary models and of the amount of convective core overshooting which occurs in stars.


\section*{Acknowledgements}

Thanks are due to Phil Dufton, Danny Lennon and Kim Venn for their generosity in providing spectra of V621\,Per, to Jens Viggo Clausen for providing {\sc sbop} and {\sc ebop}, and to Danny Lennon for discussions. We would also like to thank the referee for important suggestions, including the comparison of models and observations in the Hertzsprung-Russell diagram.

The following Starlink packages have been used: {\sc convert}, {\sc kappa}, {\sc figaro}, {\sc autophotom} and {\sc slalib}. The following internet-based resources were used in research for this paper: the WEBDA open cluster database; the ESO Digitized Sky Survey; the NASA Astrophysics Data System; the SIMBAD database operated at CDS, Strasbourg, France; the VizieR service operated at CDS, Strasbourg, France; and the ar$\chi$iv scientific paper preprint service operated by Cornell University. 

JS acknowledges financial support from PPARC in the form of a postgraduate studentship. The authors acknowledge the data analysis facilities provided by the Starlink Project which is run by CCLRC on behalf of PPARC.  SZ wishes to acknowledge support by the European RTN ``The Origin of Planetary Systems'' (PLANETS, contract number HPRN-CT-2002-00308) in the form of a fellowship. 

This paper is based on observations made with the Isaac Newton Telescope operated on the island of La Palma by the Isaac Newton Group in the Spanish Observatorio del Roque de los Muchachos of the Instituto de Astrofis\'\i ca de Canarias.

This publication makes use of data products from the Two Micron All Sky Survey, which is a joint project of the University of Massachusetts and the Infrared Processing and Analysis Center/California Institute of Technology, funded by the National Aeronautics and Space Administration and the National Science Foundation.


\end{document}